# Novel Design of Biplanar Electrodes in a Multiwell Plate for Transepithelial Electrical Resistance Measurement in 3D Cell Cultures

Georges Dubourg[a], Divyasree Prabhakaran[b], Harry Dawson[b], Vasa Radonic[a], Sara Joksović [a], Jovana Stanojev[a], Antoni Homs Corbera[b]

[a] University of Novi Sad, Biosense Institute, Dr Zorana Đinđića 1,21000 Novi Sad, Serbia

[b] Innovation Unit, Cherry Biotech SAS, Paris, 93100, France

Highlights:

- Multiwell plate with integrated electrodes enables advanced impedance spectroscopy in 3D cell cultures.
- The device offers high-throughput capabilities and optical access for microscopy analysis in 3D cell cultures.
- TEER values for two cell lines show the effectiveness of the developed methodology in assessing barrier function.

## Abstract

Recent advances in microphysiological systems have underscored the need for novel sensing and monitoring systems specifically designed for three-dimensional (3D) cell culture. In this article, an original architecture of a cell-culture multiwell plate embedding an impedance spectroscopy monitoring system is presented alongside a fast and straightforward fabrication process that allows scale-up production for further industrial exploitation. Our proposed approach involves a biplanar electrode configuration, enabling real-time and high-throughput impedance spectroscopy measurements through a 3D biological construct. This configuration not only allows for non-invasive electrical characterization of a 3D tissue layer, but also offers the capability to thoroughly analyze and understand the characteristics of any type of biological barrier and 3D tissue. The proof of concept for the developed system has been effectively validated for measuring impedance across both endothelial and mouse myoblast monolayers cultured within a 3D collagen biomimicking matrix. Moreover, transepithelial electrical resistance values, which vary with cell type and duration of cell culture, were effectively determined from the impedance spectra. The obtained results showed that the proposed device allows the monitoring of tight junction and cell layer formation over time without inducing alterations to the cells.

1. Introduction

Significant progress has been made in three-dimensional (3D) cell culture with the advancement of microphysiological systems (MPS), organ-on-chip technology, and 3D organoids. This progress offers great promise for revolutionizing the drug development process in the pharmaceutical industry as it can overcome the limitations of the traditional 2D cell cultures by mimicking the complex 3D structure and microenvironment of tissues in the human body.
This progress has been complemented by the development of advanced sensing and monitoring systems, which are crucial for studying the dynamic behavior of cells within 3D tissue models [1], [2]. Prospective sensors ought to possess the capability to effectively monitor 3D cell culture systems in real-time, employing non-invasive approaches [3]. Among potential monitoring systems such as biosensors or microscopy, electrochemical impedance spectroscopy (EIS) is a widely accepted non-destructive method [4], [5]. In particular, by analyzing the impedance spectra, valuable insights can be gained regarding the morphological alterations in cells and the tissue's cellular organization. Exploiting the impedance spectrum also enables the measurement of transepithelial electrical resistance (TEER) across a broader frequency range, thereby enhancing the amount of information obtainable from electric measurements. TEER measurement is a label-free and effective method for quantifying the barrier integrity of cellular monolayers in vitro [6], [7]. To date, advances in microfluidic-based organ-on-chip allow the fabrication of sophisticated devices integrating electrodes to provide the measurement of TEER of various epithelial and endothelial barriers in culture, including lung [8], gut, [9], [10] skin, [11], heart [12] and the blood-brain-barrier [13]. Such sensors usually consist of two pairs of facing electrodes separated by a cell culture layer or biological construct allowing the impedance measurement inside a biological construct that can provide additional information such as tissue barrier integrity [14], [15].

However, the lack of standardization in microfluidics heavily constrains the commercialization of organ-on-chip based on microfluidic technologies [16], [17]. Indeed, without standardized protocols and guidelines, it becomes more challenging for companies to develop products that meet the industrial needs across the healthcare and pharmaceutical sectors [18].

Therefore, the utilization of standard multiwell formats for constructing MPS in standard multiwell plates is an increasingly promising approach as they offer reproducibility and dimension uniformity which reduce inter-well and inter-plate variations. [19-21]. Their use simplifies experimental procedures, enabling protocol adherence and result comparison across different experiments or labs. Several devices for impedance spectroscopy measurement have been proposed in the literature or can be found in the market [21- 24]. However, most of the examples of multiwell plates for TEER measurement involve the use of external chopsticks or pin electrodes that are placed into the well during cell culture studies [25-28]. Those approaches can complicate the experiments in traditional well plates and lead to issues such as variability and technical challenges due to factors like electrode positioning and impedance fluctuations. Incorporating and immobilizing TEER recording electrodes directly within the cell culture well, positioned in close proximity to the cell monolayer, will minimize signal noise arising from electrode movement.

In this work, an original approach of biplanar electrodes positioned across two levels of a cell culture well for conducting impedance spectroscopy is proposed. The proposed approach is simple and easily scalable for further industrial production and does not require the introduction of external electrodes during cell culture experiments. This technological process is based on a manifold assembly combined with laser micromachining and micro stencil technique for the design of the electrodes. As proof-of-concept, we demonstrated the applicability of our device for measuring impedance spectroscopy and TEER by developing two 3D biological models: an endothelial barrier model and a myoblast cell monolayer cultured on a 3D collagen biomimicking matrix.

This instrument presents a significant advance due to its high throughput capabilities, user-friendliness, and capacity to investigate cellular behavior and barrier integrity using standard cell culture protocols. We anticipate widespread adoption of this technology within academic and pharmaceutical research communities for routine experimentation and analysis.

## 2. Material and Method

### 2.1. Device Design

Figure 1a provides a schematic representation of a single cell culture well. The multiwell plate incorporates biplanar electrodes, comprising two sets of electrodes positioned at two different planes within the well: the bottom and suspended electrodes. The bottom electrodes are located on the upper side of the bottom layer, while the suspended electrodes are positioned on the lower side of a suspended layer, at the periphery of the well. The two sets of electrodes are separated by an intermediary layer, thus generating a gap between the suspended and bottom electrodes that allows for the cultivation of a 3D biological construct. It should be noted that the extent of this gap can be modified by adjusting the thickness of the intermediary layer. The upper part is a perforated layer that defines the structure of the well.

In this approach, endothelial cells are cultured directly on a collagen matrix deposited onto the bottom electrodes, rather than on a separate membrane. Traditional TEER-enabled cell culture plates often rely on permeable inserts, which introduce complexity, potential artifacts, restricted access to the cell layer, and possible effects on cell phenotype. In contrast, membrane-free devices allow cells to grow directly on an optimized substrate, eliminating artifacts from membrane properties like thickness, pore and density [28], [29]. As a result, the measured TEER values more accurately reflect the true resistance of the cell monolayer without artifacts from the underlying membrane. Additionally, while membrane-based systems require precise electrode alignment, our approach simplifies handling by integrating electrodes, thereby reducing misalignment errors. The key advantages of these advanced systems include improved reproducibility by eliminating variability in membrane quality and enhanced imaging and monitoring, as the absence of additional membrane allows for clear visualization of the cell layer.

A simplified equivalent model, adapted from previous studies, is shown in Figure 1b [30], [31]. In this model, the endothelial cell layer is represented as a parallel electrical circuit, where $R_{TEER}$ corresponds to the trans-epithelial electrical resistance, and $C_m$ represents the cell membrane capacitance. This endothelial layer is modeled in series with $R_{media}$, the resistance of the culture medium. It is important to note that the influence of the double-layer capacitance at the electrode-culture medium interface has been excluded from the model. This simplification is justified by the implementation of a four-terminal measurement approach, which effectively minimizes the influence of electrode polarization, thereby enhancing the sensitivity of the measurements to the electrical properties of the cell layer [32].

The impedance spectroscopy was measured under an AC (alternative current) current using the impedance analyser (MFIA 5MHz, Zurich Instruments AG, Zurich, Switzerland), and to minimize measurement errors, a four-terminal method was employed consisting of two exciting electrodes and two measuring electrodes as shown in Figure 1c. To be more precise, a pair of diagonally opposite electrodes are utilized as exciting electrodes to apply an AC signal, while the other two electrodes are employed for measuring the impedance within the cell culture area. This electrode configuration has been specifically designed to ensure maximum electric field concentration in the cell culture area as shown in previous works [14], [33].

The method for calculating TEER, as outlined in [34], entails subtracting the average impedance values observed within the low frequency range (100 Hz to 1 kHz) from the corresponding blank values (impedance measurements obtained in the absence of cells).

Subsequently, this resulting value is normalized relative to the surface area of the cell culture well, typically standardized at 2 cm².

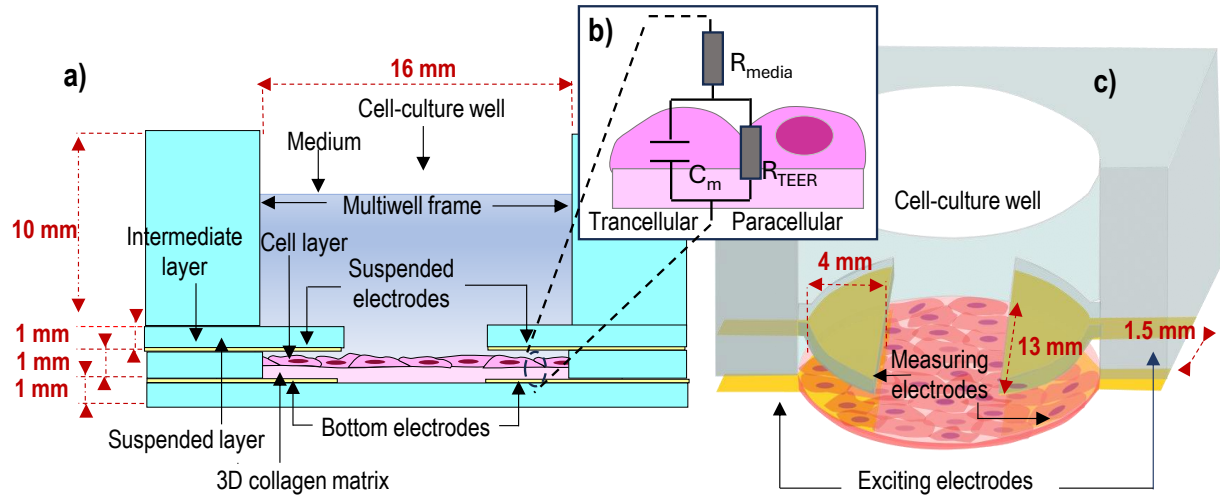

**Figure 1.** Schematic illustration of one individual cell culture well including both **(a)** cross-sectional, **(b)** Equivalent electrical circuit diagram and **(c)** three-dimensional views.

2.2. Fabrication of the biplanar electrodes in a multiwell plate

Keeping closely in mind the further industrial exploitation, our objective was to create a straightforward, cost-effective and potentially scalable technological process for incorporating biplanar electrodes into a multiwell plate. This initiative sought to preserve the attractiveness of this standard as a low-cost device while offering enhanced functionality. The fabrication process of the devices proposed in this work is not only rapid but also eliminates the need for expensive semiconductor equipment and high-temperature steps typically associated with silicon manufacturing. The schematic representation of the fabrication process for the multiwell plate is depicted in Figure 2. The multiwell manufacturing process is based on a multi-layer assembly technique which involves the parallel processing of multiple layers. As shown in Figure 1a, the multiwell plate is composed of one bottom, layer, one intermediate layer, one suspended layer and one multiwell frame that are stacked on top of one another. Therefore, to initiate the procedure, a 1 mm-thick polycarbonate (PC) layer was precisely cut using a $CO_2$ laser engraving cutting machine (MBL 4040) to create the bottom, the supporting, and the intermediate layers as shown in Figure 2a. In parallel, a 3M™ GPT-020C double-sided adhesive layer used for the connection of the PC layers was also cut into three layers using the $CO_2$ laser as shown in Figure 2b. Note that, the laser parameters have been adjusted based on the material used in the process, as shown in Table S1.

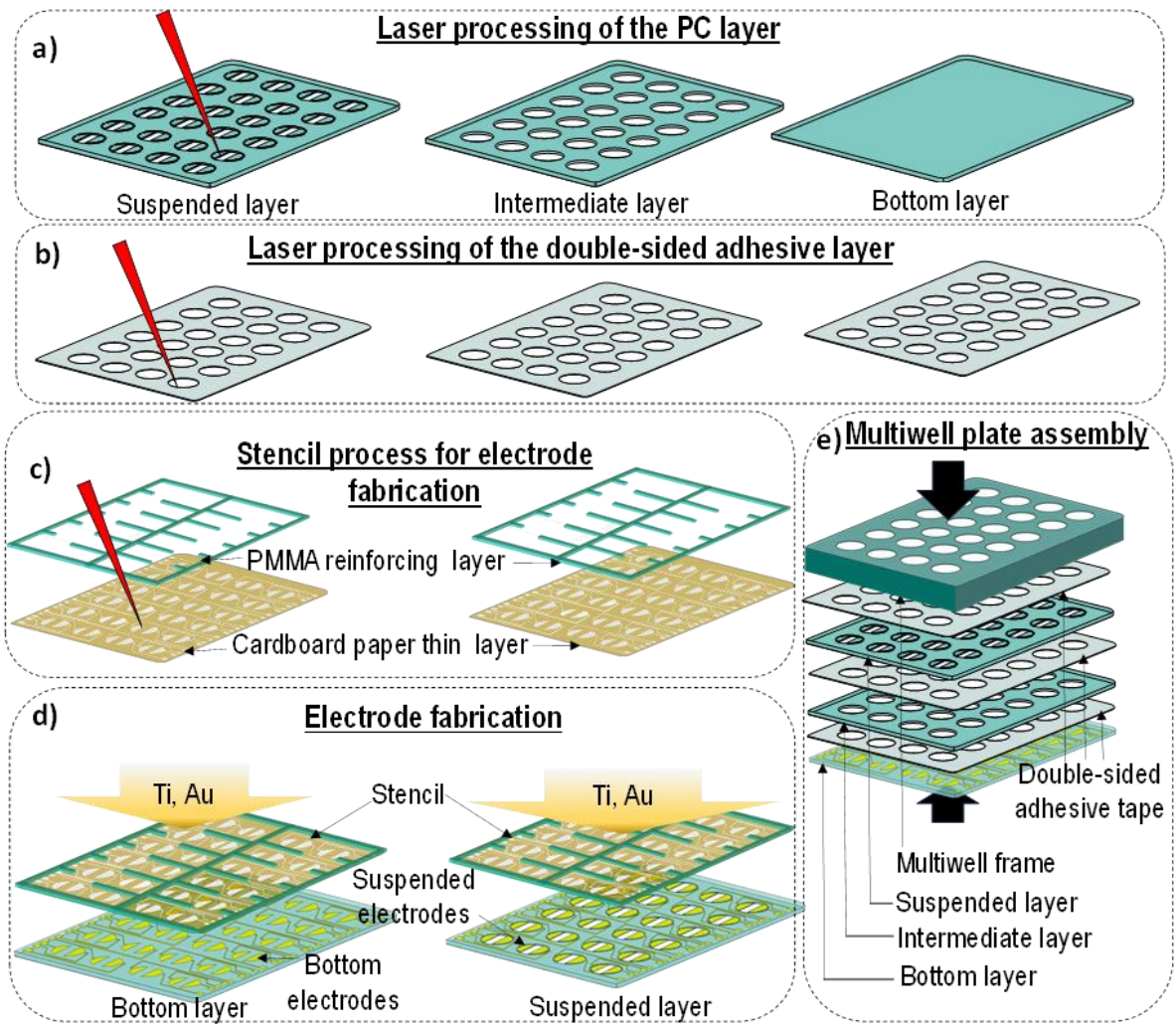

**Figure 2.** Process sequence of the multiwell plate: (**a**) Laser processing of the different multiwell layers (**b**) Laser processing of the double-sided adhesive tape, (**c**) Fabrication of the microstencil, (**d**) Deposition of the gold layer, (**e**) Assembly of the multiwell plate.

The subsequent aspect to be contemplated involves the design of gold electrodes on both the bottom and suspended layers. This work presents a versatile approach for producing the electrodes that eliminates the need for photo-masking and chemical etching steps typically used in gold patterning processes like lift-off or chemical etching. This is especially important because the layers may not withstand the solvents used in such processes. The stencil utilized in this procedure comprises two layers: a thin layer employed for defining the electrode design, and a reinforcing layer employed to guarantee an effective mechanical contact between the thin layer and the PC layers, thereby minimizing the occurrence of shadow effects. The thin layer was made of a sheet of cardboard paper, while the reinforcing layer was made of a 2 mm thickness of Poly (methyl methacrylate) (PMMA). Both layers underwent patterning through the utilization of a $CO_2$ laser (Figure 2c). Afterward, a thin adhesive layer of titanium (50 nm) and a gold layer (150 nm) were subsequently deposited on the surface of the bottom and supporting layers by thermal evaporation (through the previously fabricated stencil as shown in Figure 2d. In parallel, to create a multiwell frame designed to be positioned on the suspended layer, a 1cm-thick PC underwent precision machining using a CNC (Computer Numerical Control) machine from Protolabs (Poland). In the final processing step, the complete multiwell plate assembly involved stacking all the processed layers together using the previously patterned double-sided adhesive tape as shown in Figure 2e. Here, the double-sided adhesive tape provides a strong and reliable bond between each layer of the multiwell plate structure, preventing any potential shifting, detachment, or leakage during cell culture use. After the assembly of the multiwell plate, it was left overnight with a weight placed on top of the multiwell plate to apply consistent pressure and facilitate optimal bonding between the layers.

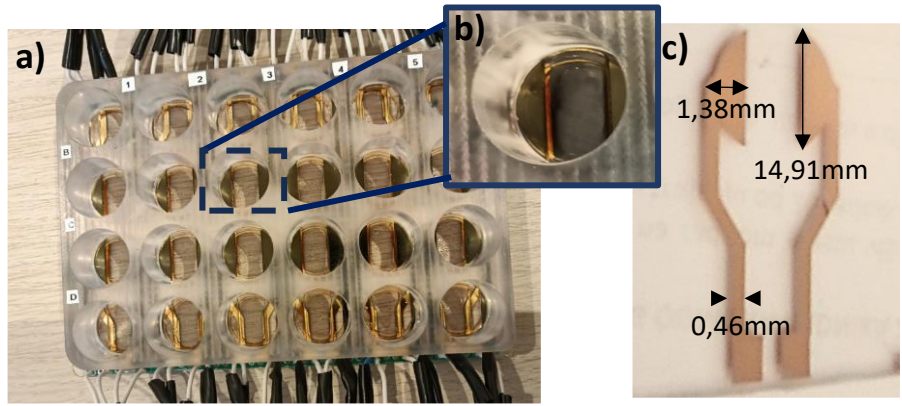

**Figure 3**. (**a**) Picture of a complete 24 multiwell plate. (**b**) Zoom in a single cell culture well (**c**) Optical image of one pair of electrodes.

Figure 3 represents the completed device, which includes a multiwell plate with connecting wires (Figure 3a), a detailed view of an individual cell culture well (Figure 3b), and Figure 3c representing the bottom electrodes with geometric characterization conducted utilizing a Stylus Profilometer, Bruker DektakXTbrucker (Profile measurements shown in Figure S1). The image demonstrates that the surface of the electrode structures exhibits spatial resolution that aligns closely with the dimensions specified in the initial design. The consistent and well-defined features observed on the electrodes highlight the precision and effectiveness of the manufacturing process.

2.3. Preparation for cell-culture in the multiwell plate

Initially, the multiwell plate was sterilized using a solution consisting of ethanol (70%) diluted in deionized (DI) water (30%). This sterilization process effectively eliminates any potential contaminants present in the multiwell plate. Subsequently, the plate was left to air dry for a period of 12 h under the cell culture hood to ensure complete evaporation of any remaining moisture and solvent. Next, the multiwell plate was coated with RatCol® rat tail collagen specifically designed for 3D hydrogels (5153, Advanced Biomatrix, Sigma Aldrich, France). The application of RatCol® rat tail collagen onto the surface of the multiwell plate provides a suitable substrate for cell adhesion and growth in a 3D environment. The coating hydrogel was prepared by mixing the collagen, 4.1 mg/mL, and neutralization solution (5155, Advanced Biomatrix, Sigma Aldrich, France) in a 9 to 1 proportion to achieve a final concentration of 3,69 mg/mL.

In this study, two distinct cell types were employed to validate the multiwell plate for TEER characterization. Commercially available Human Umbilical Vein Endothelial Cells (HUVEC) were selected as one of the cell types due to the formation of tight junctions. The use of HUVECs in this context allows for the assessment of TEER values in a cell population where tight junctions are expected to be present, providing a reference for barrier function.

Conversely, a myoblast cell line C2C12 (CRL-1772, ATCC, France) was utilized as the second cell type for which no tight junctions are anticipated. This choice enables the evaluation of TEER values in a cell population lacking tight junctions, serving as a control group to compare barrier properties in the absence of these junctions. By incorporating both HUVECs and myoblasts in the TEER characterization process, this study aims to validate the multiwell plate for assessing barrier function under different cellular conditions and junctional complexities. Both cell lines cells were washed with 15 mL of 4-(2-hydroxyethyl)-1-piperazineethanesulfonic acid buffer (HEPES, CC5024, Lonza, Colmar, France). Then, 7 mL of trypsin 0.025% Ethylenediaminetetraacetic acid (EDTA) solution (CC-5012, Lonza, Colmar, France) was added while keeping the flask in the incubator for 4 min to detach the cells. Immediately after 15 mL of Trypsin Neutrilization solution (TNS, CC-5002, Lonza, Colmar, France) was added and cells with the medium were collected in a 50 mL tube. After a 5 min centrifugation at 200G, were resuspended in EBM to reach the right cell density in the wells of the TEER plate which was verified with Burker chamber counting. The HUVEC solution was placed in each well of the plate and cells were left on the incubator to adhere to the substrate and to reach confluence. Media was then replaced by Endothelial Cell Growth Basal Medium-2 (EBM2, CC-3162, Lonza, Colmar, France). In the meantime, C2C12 cells were resuspended in 1 mL of Dulbecco's Modified Eagle Medium (DMEM, Thermofischer, France) with 20% fetal bovine serum (FBS, Merk, France) and 1% Penstrep, D20. The C2C12 solution was placed in the multiwell plate and cells were left on the incubator to adhere to the substrate and reach the confluence.

### 2.4. Microscopy characterization

Microscope images were captured after each electrical measurement to visually track the progression and proliferation of cells utilizing a bright-field microscope (Leica DMIL LED Fluo Microscope). The live/dead staining procedure was conducted as follows: the live/dead reagents (Thermofisher kit LIVE/DEAD), consisting of 5 μL calcein-AM and 20 μL ethidium homodimer (ratio of 1 to 4) mixed in 10 mL phosphate-buffered saline (PBS). Subsequently, the cells cultured in the cell culture wells were subjected to staining with the prepared live/dead reagents. Following the staining process, images of the stained cells were examined using a fluorescence microscope (Leica Thunder Imager Live Cell and 3D assay confocal fluorescent microscope).

## 3. Results and discussion

### 3.1. Optical characterization of cell viability in the multiwell plate

Initially, cell adhesion and proliferation on the collagen layer within the fabricated multiwell plate during electrical measurements were qualitatively assessed using standard phase contrast microscopy. This evaluation aimed to verify the cells' ability to adhere to the collagen layer and proliferate seamlessly under the provided cell culture conditions, while ensuring that the electrical measurements do not induce any alterations to the cells. One of the main advantages of this system is highlighted here. Observing cells in a standard multiwell plate can be challenging due to their opacity, irregular surfaces, and the thickness created by collagen membranes, as well as the obstructed field of view caused by electrodes. These obstacles can hinder the ability to accurately study cells in real-time through microscopy. Figure 4 illustrates optical images acquired during impedance measurements at different time points, depicting HUVEC at 18 h, 24 h, and 48 h (Figures 4a, b, c), alongside C2C12 cells at 18 h, 36 h, and 48h (Figures 4e, f, g). The pictures indicate that both cell types successfully adhered to the collagen surface and proliferated over time to create a monolayer of cells. Figure 4b shows that HUVECs achieved full confluence within 24 hours, consistent with findings from other studies reporting a confluent state at the same time point [35-37]. This observation highlights the progressive formation and stabilization of a cohesive monolayer, with the barrier properties of the cell layer achieving a steady state at this time point. In contrast, Figure 4 indicates that C2C12 cells reached full confluence later, beyond 24 hours. This suggests that the formation of a cohesive and functional monolayer in C2C12 cells requires more time compared to HUVECs. This difference can be attributed to intrinsic variations in the cellular properties and growth dynamics of endothelial cells (HUVECs) and myoblast cells (C2C12).

Furthermore, based on the microscope images presented in Figures 4d and 4h depicting control wells for HUVEC and C2C12 in which electrical measurements were not conducted, it is apparent that there are no discernible differences between the control wells and other wells. This observation indicates that the electrical measurements did not exert a significant influence on the cell culture under examination.

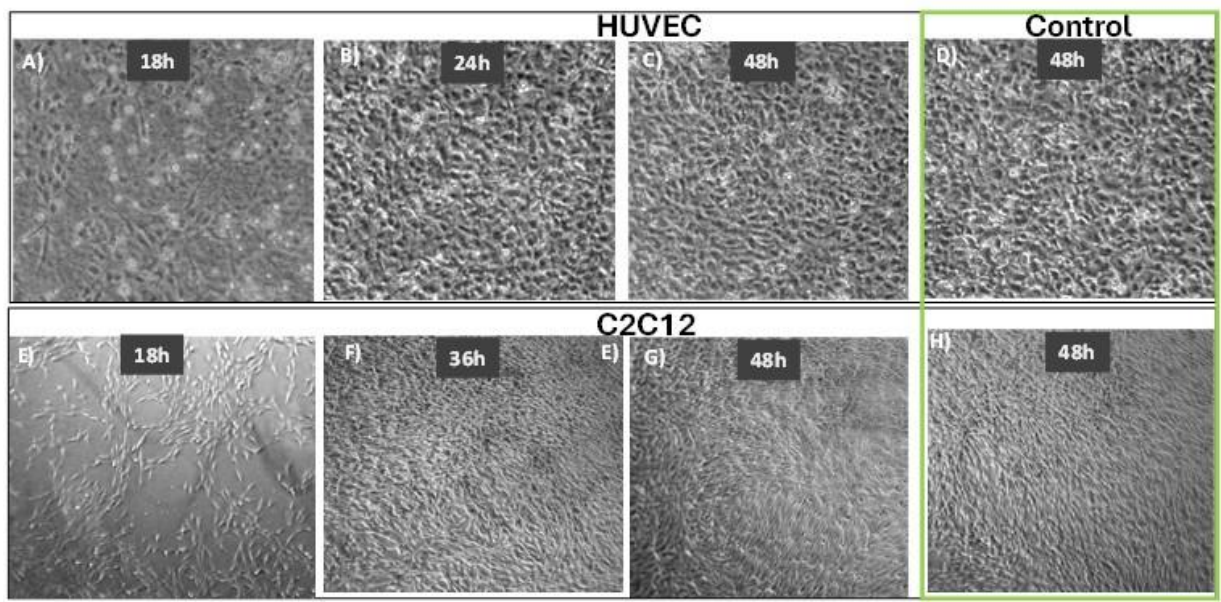

**Figure 4**. (**a**), (**b**), (**c**) Microscope images of HUVEC cells at 18 h, 24 h and 48 h respectively. (**e**), (**f**), (**g**) Microscope images of C2C12 cells at 18 h, 36 h and 48 h respectively. (**d**) and (**h**) microscope images of control wells where no measurements were conducted. (10x magnification)

In addition, live/dead assays were performed to confirm the viability of the HUVEC and C2C12 cells after 48 h. The fluorescence analysis results, depicted in Figures 5a and 5b for HUVEC and C2C12 cells, respectively, indicate the proportion of live/dead cells within the multiwell plate. The presence of a green signal in both cell types indicates excellent cell viability, exceeding 95%, when cultured on the collagen-coated surface within the multiwell plate. This observation underscores the favorable growth conditions and cellular health maintained during the experimental timeframe.

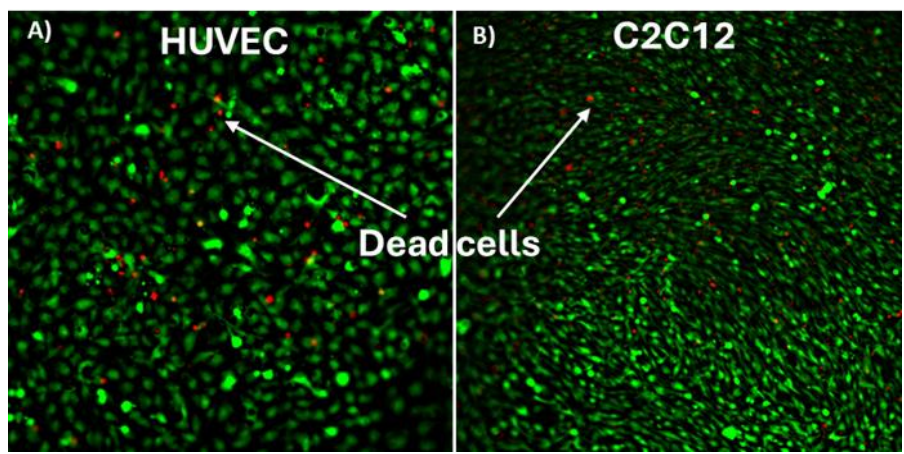

**Figure 5.** Fluorescence micrographs of **(a)** HUVEC and **(b)** C2C12 following immunofluorescence live-dead staining (10x magnification).

### 3.2. Electrochemical impedance spectroscopy in 3D cell culture conditions.

Before utilizing the multiwell plate for 3D cell culture, the amplitude of the applied voltage was optimized by measuring impedance across a frequency range of 100 Hz to 1 MHz at various voltage levels (10 mV, 30 mV, 100 mV, 300 mV, and 1 V) under different media conditions, including EBM, EBM2 and DMEM. The results, presented in Figure 6a. b and c respectively, reveal that the impedance curves could not be measured correctly until reaching voltages over 300 mV for all media, beyond which the measurements remained consistent. Therefore, an amplitude voltage of 300 mV was selected for subsequent measurements, as it ensures stable and reliable impedance readings while minimizing the applied voltage to avoid any potential effects on the cells. Figure 6d illustrates the impedance curves for EBM, EBM-2, and DMEM, measured under an applied voltage of 300 mV. The impedance curves for EBM and EBM-2 are observed to overlap, which is expected given their nearly identical compositions. This overlap confirms the consistency of their electrical properties due to the minimal differences in their formulations. In comparison, the impedance curve for DMEM is lower than those of EBM and EBM-2. This difference can be attributed to the distinct composition of DMEM, which is designed for a broader range of cell types and contains a different balance of salts, nutrients, and supplements. These compositional differences influence the medium's ionic strength, conductivity, and, consequently, its impedance profile. Additional measurements performed with extra wells and Collagen/EBM, shown in Figures S2 and S3, confirmed these trends.

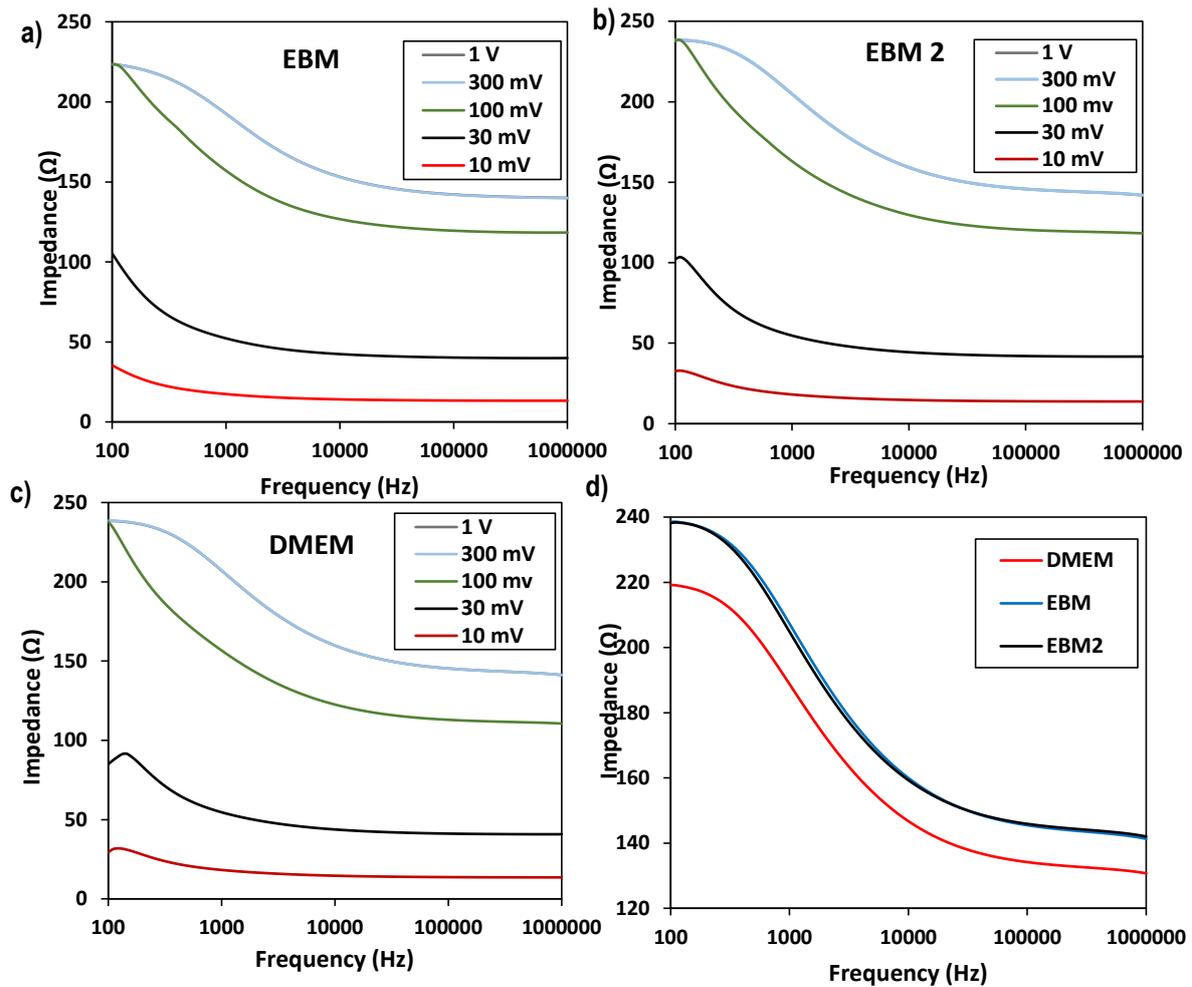

**Figure 6.** Impedance measurements performed at different applied voltages using: **(a)** EBM, **(b)** EBM2, and **(c)** DMEM. **(d)** Impedance curve for DMEM, EBM, and EBM2 measured at 300mV.

To assess the effectiveness of the multiwell plate for impedance measurements, we observed the changes in impedance corresponding to cellular growth and proliferation within the collagen layer over time. For each cell type, impedance spectroscopy was subsequently measured in four distinct cell culture wells, each containing a 200 μm-thin collagen layer, seeded with cells and filled with culture medium.

Once cells were attached to the multi well-plate device, impedance measurements were recorded at several time intervals. The cells on the electrode-based devices were then observed under a microscope to assess the cell culture. The data collected from impedance spectroscopy measurements, in a frequency range of 100 Hz to 1 MHz, is presented in Figure 7.

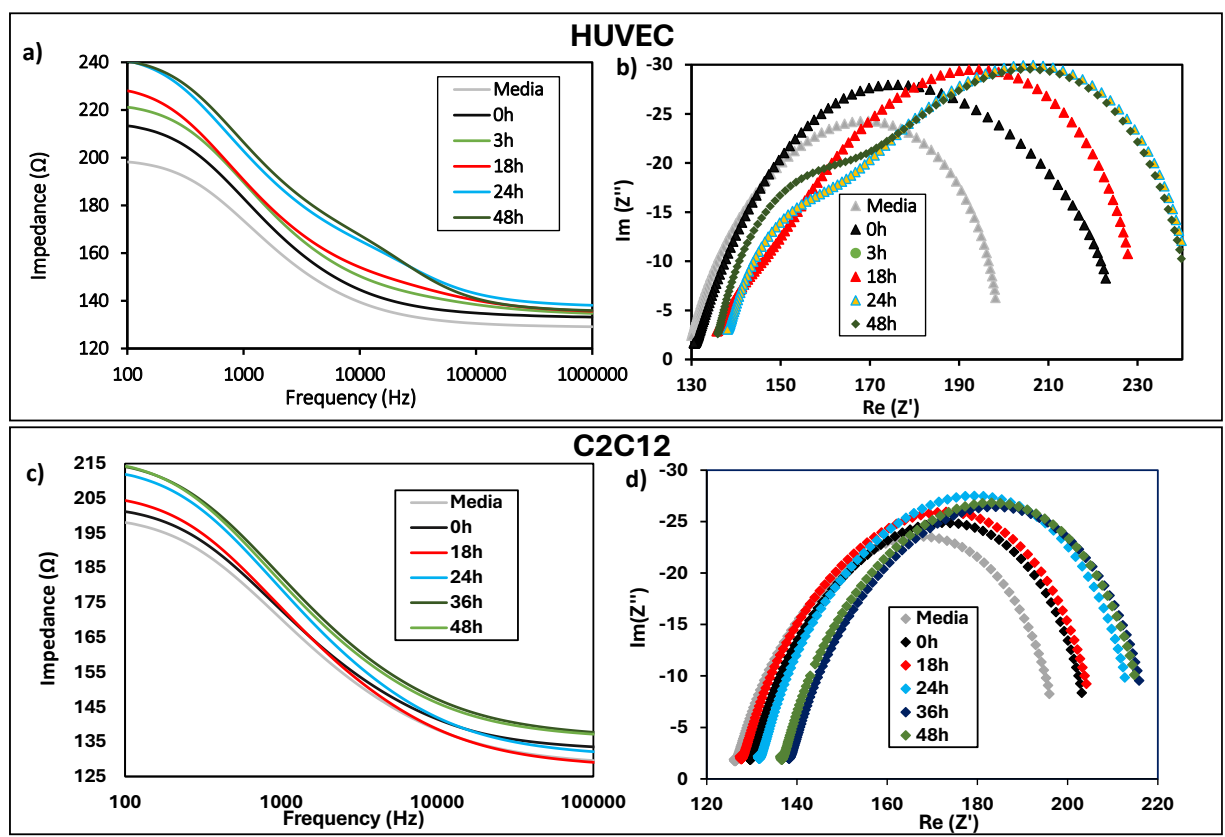

**Figure 7. (a)** and **(c)** Impedance measurements conducted at different time points across a cell culture well in the presence of HUVEC and C2C12 cells, respectively. **(b)** and **(d)** Nyquist plots of impedance analysis of HUVEC and C2C12.

In both cases, the impedance values increase in the positive direction, indicating the presence and growth of cells as well as the development of a cell layer on the bottom electrode surface. Nyquist impedance plots (-Im(Z″) versus Re(Z′)) were generated to illustrate this relationship (Figures 7b and 7d). The data reveal that the radius of the Nyquist semi-circle typically increases over time until day 1, reflecting the progression of cell growth. By day 2, the radius remains unchanged, signifying that HUVEC cells have reached complete confluence by day 1. From the more detailed analysis in Figure 7a, we observe an impedance increase from 195 Ω to 210 Ω following cell seeding, compared to the baseline impedance of the culture medium without cells. This rise is attributed to changes in the medium's conductivity due to the presence of cells and their initial attachment to the collagen layer. As HUVEC cells interact with the medium, they alter its ionic composition, reducing conductivity and increasing impedance, which reflects their impact on the solution's electrical properties. Subsequently, as the cells adhere and form a cohesive monolayer on the collagen layer, the impedance at lower frequencies (e.g. 100 Hz) rises significantly, reaching approximately 230 Ω after 18 hours and further increasing to 240 Ω after 24 hours. This corresponds to an overall impedance increase of about 45 Ω.

At low frequencies, the transcellular capacitor becomes charged, effectively blocking current flow through the transcellular pathway. Consequently, low-frequency alternating current is primarily redirected through the trans-epithelial electrical resistance (TEER), which reflects the resistance of the paracellular pathway, including contributions from tight junctions and intercellular spaces [34], [38].

In contrast, at higher frequencies (e.g. above 1 kHz), the impedance curve exhibits a different behavior, with only minimal changes observed as the frequency approaches 1 MHz. At these higher frequencies, the transcellular capacitor becomes increasingly conductive, allowing current to flow through the transcellular pathway. As the frequency increases, the capacitive contributions of cell membranes and tight junctions diminish, resulting in a reduction in overall impedance. This frequency-dependent behavior explains the convergence of the impedance curve toward the baseline conductivity of the culture medium at higher frequencies, where the cellular barrier's influence becomes negligible [34], [38].

In Figure 7d, the behavior of C2C12 cells is shown. The impedance changes are less pronounced compared to HUVEC cells, with a maximum increase of 18 Ω at 100 Hz and 10 Ω at 1 MHz after 36 hours. This is consistent with the fact that C2C12 cells do not form tight junctions. Therefore, the impedance increase in this case is primarily due to cell proliferation rather than the formation of a cohesive monolayer with tight junctions.

### 3.3. Determination and analysis of TEER values

Afterwards, the TEER values were determined from the measured impedance spectra in a range of 100 Hz to 1025 Hz (Impedance spectra of the wells are shown in Figure S4).

In Figure 8, the TEER values for four different cell culture wells are illustrated. Specifically, Figure 8a displays the TEER values calculated for HUVEC cells, while Figure 8b showcases the TEER determined for C2C12 cells. The data indicates that the TEER values for both types of cells steadily increased before reaching a plateau, with a slight decrease observed in the case of endothelial cells after 24h. The plateau phase was reached at 24 h for HUVEC cells, whereas for C2C12 cells it occurred after 36 h.

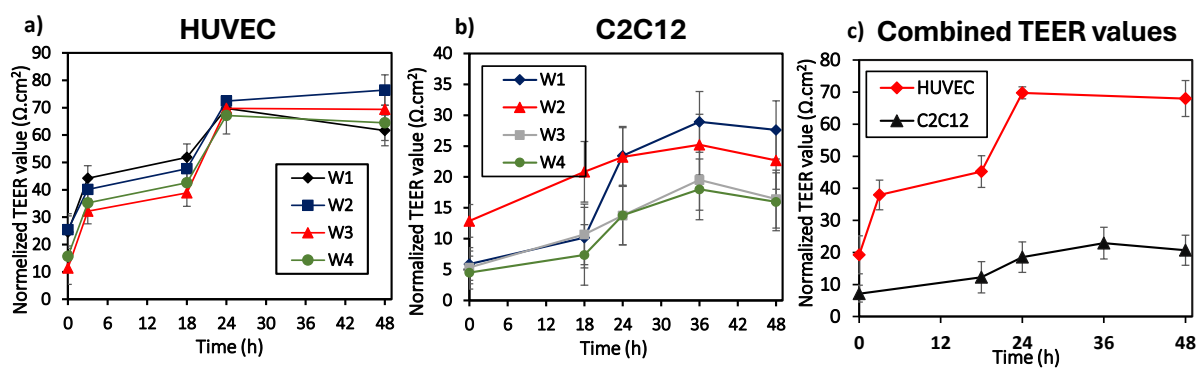

**Figure 8.** Transepithelial electrical resistance (TEER) versus time for HUVEC **(a)** and C2C12 **(b)** cells grown in four cell culture wells, **(c)** Combined TEER values of the 4 individual wells for both cell lines. (Error bars represent the standard deviation).

Figure 8a shows that the TEER values for HUVEC cells at confluence ranged from 50 to 80 Ω·cm², aligning with reported values from various commercial TEER measurement systems. The table below provides a comparative summary of TEER value ranges for HUVECs at the confluent state, as documented in previous studies employing commercial setups.

**TABLE 1**: Summary of TEER values for HUVECs at confluent state using various commercial setups

| TEER (Ω·cm²) | Cell-culture device | Measurement set up | Ref. |
|---|---|---|---|
| 5-100 | Transwell | EVOM (Ohm's Law) | [36-37], [39-42] |
| 15-100 | Transwell | Millicell ERS-2 (Ohm's Law) | [43], [44] |
| 400-500 | Transwell | Millicell ERS-2 (Ohm's Law) | [36], [45] [46] |
| 350-500 | Transwell | EVOM (Ohm's Law) | [47] |
| 5-30 | OrganoPlate | OrganoTEER (Impedance spectroscopy) | [48], [32] |
| 5-10 | Transwell | cellZscope (Impedance spectroscopy) | [49] |
| 60-80 | Multiwell plate | Impedance spectroscopy | This work |

Table 1 shows that the TEER values for a confluent layer of HUVECs in our study are consistent with those reported in the literature using commercial TEER measurement instruments. Most of these studies employ a transwell insert, a membrane on which HUVECs are cultured. For example, studies using the EVOM system (World Precision Instruments, Sarasota, Florida), a volt-ohmmeter, have reported TEER values ranging from 5 to 100 Ω·cm² [36-37], [39-42]. Similarly, the Millicell ERS-2 system (Millipore, Darmstadt, Germany), another volt-ohmmeter, has documented values between 15 and 100 Ω·cm² [43], [44]. Impedance-based measurement devices, such as the OrganoTEER (MIMETAS B.V., Oegstgeest, The Netherlands), have recorded TEER values ranging from 5 to 30 Ω·cm² in the OrganoPlate (MIMETAS B.V., Oegstgeest, The Netherlands), an organ-on-chip multiwell plate [48], [32]. Additionally, the cellZscope (NanoAnalytics, Münster, Germany), when used with a transwell insert, has reported TEER values between 5 and 10 Ω·cm² [49]. In contrast, some investigations using the EVOM and Millicell ERS-2 systems recorded substantially higher TEER values, ranging from 300 to 500 Ω·cm² [36], [44-46]. This variability in TEER measurements across studies can be attributed to factors such as the technical specifications of measurement devices (e.g., electrode configuration and positioning) and the characteristics of the membrane or substrate (e.g., material composition, thickness, pore size, and density). Notwithstanding these differences, the TEER values reported in this study exhibit good concordance with those obtained using widely recognized commercial systems underscoring the reliability and robustness of the proposed experimental setup [50-52].

In Figure 8c, it can be observed that the TEER values of HUVEC cells are higher compared to those of C2C12 cells. This difference is mainly due to fundamental differences in the biological roles and structural properties of these two cell types. C2C12 cells, derived from mouse skeletal muscle, are primarily designed for muscle development and differentiation rather than the formation of tight barriers. Consequently, they exhibit low expression levels of tight junction proteins, such as claudins, occludins, and ZO-1, which are crucial for establishing high electrical resistance [53], [54]. In contrast, HUVEC cells, which are human umbilical vein endothelial cells, are specialized to maintain vascular integrity and regulate selective permeability in blood vessels. These endothelial cells naturally form cohesive monolayers with strong tight junctions, resulting in significantly higher TEER values. The observed variation reflects the distinct biological roles of these cell types. HUVEC cells, as endothelial cells, form tight junctions that restrict the passage of ions and molecules through the paracellular pathway, thereby contributing to the elevated electrical resistance of their monolayer. On the other hand, C2C12 cells, which do not typically form tight junctions, exhibit lower TEER values. The slight increase in TEER observed in the C2C12 monolayer indicates cell presence and growth but lacks the contribution from tight junction formation seen in HUVEC cells. This observation is consistent with the previous study showing that C2C12 cells consistently exhibit lower TEER values compared to endothelial and epithelial cell types, which are specialized to form tight junctions [55].

4. Conclusions

In this work, an original architecture of a multiwell plate, which incorporates multiplanar electrodes for impedance spectroscopy measurement, has been demonstrated to be compatible with cell-culture conditions and enables monitoring of TEER. This device serves as a robust toolkit for the continuous assessment of cell-culture growth and barrier tissue integrity. In addition to its compact standalone impedance meter for high-throughput TEER measurement, the unique aspects of the multiwell plate include its optical windows and four electrodes. These electrodes enable tetrapolar TEER measurements while providing a clear field of view for imaging purposes. The presented platform is applicable to any biological barrier and tissue to mimic the tissue–tissue and tissue–liquid interface. Moreover, the possibility of integrating other electrodes at several planes then opens the door for 3D electrical screening of complex 3D tissues with an increased level of complexity. Therefore, the presented platform represents a significant leap forward in biomedical research instrumentation, offering a sophisticated solution for the comprehensive study of cellular behavior and barrier integrity in vitro. The multifaceted design and adaptable functionality present an exciting avenue for further exploration and discovery within the realm of biological sciences, specifically in the development of complex 3D in-vitro models for applications such as drug development.

Acknowledgments

The work described in this article has been conducted within the projects CISTEM and PORTable STaNDOuTs that received funding from the European Union's Horizon 2020 research and innovation programme under the Marie Skłodowska–Curie Grant Agreements No. 778354 and No. 884823 respectively. In addition, this research was supported by ANTARES project that has received funding from the European Union's Horizon 2020 research and innovation programme under the grant agreements SGA-CSA. No. 739570 and FPA No. 664387. Moreover, the Transition project IDEFIX, funded by the European Innovation Council (EIC) under the Horizon Europe Grant No. 101099775, provided further backing for this research. This work was also supported by the Ministry of science, technological development and innovation of Serbia (Grant No. 2024: 451-03-66/2024-03/ 200358).